\documentclass[twoside,11pt]{article}
\usepackage{times,amsmath,amsfonts,amsthm,amssymb,eucal}
\usepackage{graphicx,ifthen}
\usepackage{wrapfig}             
\usepackage{latexsym}
\usepackage{color}
\usepackage{mathrsfs}
\usepackage[all,knot]{xy}
\xyoption{arc}
\begin{document}
\renewcommand{\baselinestretch}{1}
\def\Quote{\begin{quotation}\normalfont\small}
\def\EndQuote{\end{quotation}\rm}
\def\BigHeading{\bfseries\Large}\def\MediumHeading{\bfseries\large}
\def\bct{\begin{center}}
\def\ect{\end{center}}
\font\BigCaps=cmcsc9 scaled \magstep 1
\font\BigSlant=cmsl10    scaled \magstep 1
\def\lbk{\linebreak}
\def\Report{Random-effects Approach to Missing Data}
\def\Author{Nguyen, Zhang and Jiang}
\pagestyle{myheadings}
\markboth{\Author}{\Report}
\thispagestyle{empty}
\bct{\BigHeading A Random-Effects Approach to Generalized Linear Mixed Model Analysis of Incomplete Longitudinal Data}\\\vskip10pt
\BigCaps Thuan ${\rm Nguyen}^{1}$, Jiangshan ${\rm Zhang}^{2}$ and Jiming ${\rm Jiang}^{2}$\lbk
\BigSlant Oregon Health and Science ${\rm University}^{1}$ and University of California, ${\rm Davis}^{2}$
\ect
\Quote
\vskip-5pt

We propose a random-effects approach to missing values for generalized linear mixed model (GLMM) analysis. The method converts a GLMM with missing covariates to another GLMM without missing covariates. The standard GLMM analysis tools for longitudinal data then apply. The method applies, in particular, to the cases of linear mixed models and logistic regression. Performance of the method is evaluated empirically, and compared with alternative approaches, including the popular MICE procedure of multiple imputation. Theoretical justification of the method is given, and explained, for the patterns observed in the simulation studies. Two real-data examples from healthcare studies are discussed.

\EndQuote
\section{Introduction}
\hspace{4mm}
The missing-data problem has a long history in Statistics (e.g., Afifi and
Elashoff 1966, Hartley and Hocking 1971) that featured methodology and computational developments (e.g., Little and Rubin 2014). In this regard, a standard framework for handling the missing data, called missing-data mechanisms (MDM; Rubin 1976), is highly influential. Let $Y$ denote the complete data. The data are missing completely at random (MCAR) if the probability distribution of the missing does not depend on either the observed or the missing values in $Y$. Let $Y_{\rm obs}$ and $Y_{\rm mis}$ denote the observed and missing parts of $Y$. The data are missing at random (MAR) if the probability distribution of the missing depends only on $Y_{\rm obs}$; otherwise, the data are missing not at random (MNAR). Of course, MCAR is a stronger assumption than MAR, but even the latter is usually unverifiable, in practice.

The MDM have led to the development of multiple imputation (MI) methods,
which apply under the MCAR or MAR assumptions. For the most part, a MI
procedure consists of the following steps: (i) impute the missing values with values randomly drawn from some distributions to generate several complete cases data sets; (ii) perform the same analysis on each of the imputed complete data sets; and (iii) combine the results of the analysis based on different imputed data sets in some way. A version of MI called MICE (multiple imputation via chained equations) has been implemented in R (Van Buuren and Groothuis-Oudshoon 2011). In MICE, the imputation in step (i) is done in a way similar to the Gibbs sampler (e.g., Liu  2004).

Despite the impact of MI on the ``missing-data world'', it may be wondered if there is something conceptually simpler. To see this, let us first consider a simpler case of longitudinal data, in which there are missing covariates but no missing responses. Generalized linear mixed models (GLMM) have been widely used for longitudinal data in health and medical studies since the seminal paper of  Laird and Ware (1982). See, for example, Diggle {\it et al.} (2002), Jiang and Nguyen (2021). Suppose that, conditional on vectors of subject-specific random effects, $\alpha_{i}=(\alpha_{ik})_{1\leq k\leq d}$, $1\leq i\leq m$, responses $y_{it}$, $1\leq i\leq m, t\in T_{i}$ are independent with conditional probability density function (pdf), or probability mass function (pmf), given by
\begin{equation}
	f(y_{it}|\alpha)=\exp\left[\left(\frac{a_{it}}{\phi}\right)
	\{y_{it}\xi_{it}-b(\xi_{it})\}+c\left(y_{it},\frac{\phi}{a_{it}}
	\right)\right],\label{eq:GLMM}
\end{equation}
where $b(\cdot)$ and $c(\cdot,\cdot)$ are functions associated
with the exponential family (McCullagh and Nelder 1989, \S{2}),
$\phi$ is a dispersion parameter which is sometimes known, $a_{it}$ is a weight such that $a_{it}=1$ for un-grouped data; $a_{it}=l_{it}$ for grouped data when the group average is considered as response, and $l_{it}$ is the group size; and $a_{it}=l_{it}^{-1}$ when the sum of individual responses in the group is considered. Furthermore, $\xi_{it}$ is associated with a linear predictor,
\begin{equation}
	\eta_{it}=x_{it}'\beta+z_{it}'\alpha_{i},\label{eq:GLMM_pred}
\end{equation}
through a link function $g(\cdot)$, that is, $g(\xi_{it})=\eta_{it}$,
or $\xi_{it}=h(\eta_{it})$, where $h=g^{-1}$. Here $x_{it}$ and
$z_{it}$ are known vectors, and $\beta$ is a vector of unknown
fixed effects. In the case of a canonical link, we have
$\xi_{it}=\eta_{it}$. Finally, suppose that $\alpha_{1},\dots,
\alpha_{m}$ are independent with pdf $f_{\theta}(\cdot)$, where
$\theta$ is a vector of variance components. Let $\psi=(\beta',
\theta')'$, and $\vartheta=(\psi',\phi)$. As noted, in some cases,
such as binomial and Poisson distributions, $\phi$ is known, so $\psi$ represents the vector of all unknown parameters. Two special cases of the GLMM, which are widely used in practice, are the linear mixed models (LMM), in which case the exponential family is normal and $g(u)=u$, and logistic regression, in which the exponential family is binomial, $g(p)={\rm logit}(p)=\log\{p/(1-p)\}$, and there is no random effects [i.e., the term $z_{it}'\alpha_{i}$ is dropped in (\ref{eq:GLMM_pred})].

A defining feature of GLMM is the random effects. This actually gives us some idea on how to deal with the missing covariates. Namely, we are going to treat the missing covariates as random effects as well, after suitable centralizations. Note that the GLMM already has random effects to start with. So, at the end of the day, one still has a GLMM, albeit a different one. A big difference is that, this time, there are no missing values under the new GLMM.

An immediate question is whether one gains anything with this random-effects approach. Intuitively, it should. To see this, let us consider an extreme case, in which the covariate vector is $p$-dimensional, but for every response, there is at least one component of the covariates that is missing. In other words, on the right side of (\ref{eq:GLMM_pred}), there is a missing $x$ for every $i,t$. Clearly, the GLMM analysis based solely on complete data yields nothing. On the other hand, the random-effects approach provides an opportunity to make some progress, which is often (though not always) better than taking no action at all. Of course, this is a rather extreme case; in real life, one may not be so ``bad lucked''. Nonetheless, it suggests that the random-effects idea is at least worth exploring. In fact, our main finding in this paper is that the random-effects approach, called completed-covariates estimation (CCE), gains over the complete-data-only analysis, and in some cases over the MICE as well.

The next question is what to do if there are also missing responses. Another main finding of this paper is that, in this case, even if the missing responses are replaced by their best predictors based on the observed data, one does not have additional gains over the CCE.

In Section 2, we develop the CCE for the case of missing covariates only, and justify the efficiency gain theoretically. In Section 3, we consider the case of both missing responses and missing covariates, and show that, even in the case of LMM, a simpler special case of GLMM, one does not gain additional efficiency over CCE even if the missing responses are replaced by their best predictors. Simulation studies are carried out in Section 4 to evaluate the empirical performance of the proposed methods and their comparison with existing or alternative methods. Two real-data examples from healthcare studies, the first involving LMM analysis and the second logistic regression, are discussed in Section 5. We conclude with some discussions in Section 6.
\section{Missing covariates only}
\hspace{4mm}
In this section, we consider a special case, in which there are no missing responses but there are missing covariates. Let $x_{it}=(x_{itk})_{1\leq k\leq p}$ denote the vector of covariates supposed to be collected for $t\in T_{i}$, where $T_{i}$ is a subset of observational times with $n_{i}=|T_{i}|$ ($|S|$ denotes the cardinality of set $S$), and $\beta=(\beta_{k})_{1\leq k\leq p}$. Let $\alpha_{1},\dots,\alpha_{m}$ be independent vectors of random effects such that $\alpha_{i}\sim N(0,G)$. Furthermore, we allow an additional term in (\ref{eq:GLMM_pred}) to incorporate additional serial correction (e.g., Diggle {\it et al.} 2002), that is,
\begin{equation}
	\eta_{it}=x_{it}'\beta+z_{it}'\alpha_{i}+\epsilon_{it},\;\;t\in T_{i},\label{eq:GLMM_pred_1} 
\end{equation}
$i=1,\dots,m$, where $\epsilon_{i}=(\epsilon_{it})_{t\in T_{i}}\sim N(0,R_{i})$, $R_{i}$ being an $n_{i}\times n_{i}$ covariance matrix depending a vector $\rho$ of additional dispersion parameters. Also, let $\alpha_{i},\epsilon_{i}$, $i=1,\dots,m$ be independent.

Let $M_{it}$ denote the subset of $\{1,\dots,p\}$ so that $x_{itk}, k\in M_{it}$ are missing and $x_{itk}, k\in\{1,\dots,p\}\setminus M_{it}$ are observed. In case $M_{it}=\emptyset$, all of the $x_{itk}$'s are observed; in case $M_{it}=\{1,\dots,p\}$, all of the $x_{itk}$'s are missing. Then, (\ref{eq:GLMM_pred_1}) can be written as
\begin{eqnarray}
	\eta_{it}&=&\sum_{k\notin M_{it}}x_{itk}\beta_{k}+\sum_{k\in
		M_{it}}x_{itk}\beta_{k}+z_{it}'\alpha_{i}+\epsilon_{it}\nonumber\\
	&=&\sum_{k=1}^{p}x_{itk}1_{(k\notin M_{it})}\beta_{k}+\sum_{k=1}^{p}x_{itk}1_{(k\in M_{it})}\beta_{k}+z_{it}'\alpha_{i}+\epsilon_{it}.\label{eq:lt_LMM_1}
\end{eqnarray}

The next step is to model the second term on the right side of (\ref{eq:lt_LMM_1}). It is quite common that some of the covariates are not time-dependent. Without loss of generality, suppose that the first $p_{1}<p$ covariates are actually not time-dependent, that is, $x_{itk}=x_{ik}$, $1\leq k\leq p_{1}$, and the rest of the covariates are time-dependent. Let $M_{i}=\{1\leq k\leq p_{1}: x_{ik}$ is missing$\}$. It is easy to show that $M_{i}=M_{it}\cap\{1,\dots,p_{1}\}$. Write $M_{it,2}=M_{it}\cap\{p_{1}+1,\dots,p\}$. Then, we have
\begin{equation}
	\sum_{k=1}^{p}x_{itk}1_{(k\in M_{it})}\beta_{k}=\sum_{k\in M_{i}}x_{ik}\beta_{k}+\sum_{k\in M_{it,2}}x_{itk}\beta_{k}=\mu_{1i}+\mu_{2it},\label{eq:mis_decomp}
\end{equation}
$\mu_{1i}, \mu_{2it}$ defined in obvious ways. In view of the expression on the right side of (\ref{eq:mis_decomp}), we assume
\begin{equation}
	\mu_{1i}+\mu_{2it}=1_{(M_{i}\neq\emptyset)}\mu_{1}+1_{(M_{i}\cup M_{it,2}\neq\emptyset)}\gamma_{i}+1_{(M_{it,2}\neq\emptyset)}(\mu_{2t}+\delta_{it}),\label{eq:mc_model}
\end{equation}
where $\mu_{1}, \mu_{2t}$ are unknown fixed parameters, $\gamma_{i}$ is a subject-specific random effect, and $\delta_{it}$ is an additional random error. It is assumed that the $\gamma_{i}$'s are independent $N(0,\sigma_{\gamma}^{2})$; the $\delta_{it}$'s are independent $N(0,\sigma_{\delta}^{2})$; the $\gamma_{i}$'s and $\delta_{it}$'s are independent, and independent with the $\alpha$ and $\epsilon$. Now define $X_{1it}=[x_{itk}1_{(k\notin M_{it})}]_{1\leq k\leq p}$. Without loss of generality, let $\cup_{i=1}^{m}T_{i}=\{1,\dots,T\}$ be the times at which at least one data record is collected. Let $X_{2it*}$ be the $(T+1)\times 1$ vector, whose first component is $1_{(M_{i}\neq\emptyset)}$, $(t+1)$th component is $1_{(M_{it,2}\neq\emptyset)}$, and other components are $0$, and $\mu_{*}=(\mu_{1},\mu_{21},\dots,\mu_{2T})'$. Also let $Z_{it}'=[z_{it}',1_{(M_{i}\cup M_{it,2}\neq\emptyset)}]$, $v_{i}=(\alpha_{i}',\gamma_{i})'$, and $e_{it}=\epsilon_{it}+1_{(M_{it,2}\neq\emptyset)}\delta_{it}$. Then, combining (\ref{eq:lt_LMM_1}), (\ref{eq:mis_decomp}), and (\ref{eq:mc_model}), we can write
\begin{eqnarray}
	\eta_{it}&=&\sum_{k=1}^{p}x_{itk}1_{(k\notin M_{it})}\beta_{k}+1_{(M_{i}\neq\emptyset)}\mu_{1}+1_{(M_{it,2}\neq\emptyset)}\mu_{2t}\nonumber\\
	&&+z_{it}'\alpha_{i}+1_{(M_{i}\cup M_{it,2}\neq\emptyset)}\gamma_{i}+\epsilon_{it}+1_{(M_{it,2}\neq\emptyset)}\delta_{it}\nonumber\\
	&=&X_{1it}'\beta+X_{2it*}'\mu_{*}+Z_{it}'v_{i}+e_{it}.\label{eq:lt_LMM_2}
\end{eqnarray}
$i=1,\dots,m, t\in T_{i}$. Let $X_{2it*k}$ be the $k$th component of $X_{2it*}$, $1\leq k\leq q_{*}={\rm dim}(\mu_{*})$. It is possible that, for some $1\leq k\leq q_{*}$, $X_{2it*k}=0$ for all $i,t$. This happens, for example, if $M_{it}=\emptyset$ for all $1\leq i\leq m$. After dropping such components $k$, and the corresponding components in $\mu_{*}$, we can express $X_{2it*}'\mu_{*}$ as $X_{2it}'\mu$ such that no component of $X_{2it}$ is zero for all $i,t$. Thus, letting $X_{it}=(X_{1it}'\;X_{2it}')'$, and $b=(\beta',\mu')'$, we finally arrive at the following expression:
\begin{equation}
	\eta_{it}=X_{it}'b+Z_{it}'v_{i}+e_{it},\;\;t\in T_{i},\label{eq:GLMM_pred_2}
\end{equation}
$i=1,\dots,m$. Note that the right side of (\ref{eq:GLMM_pred_2}) is the same as the right side of (\ref{eq:GLMM_pred_1}) except with $x_{it}$, $z_{it}$ replaced by $X_{it}$, $Z_{it}$, and $\beta, \alpha_{i}, \epsilon_{it}$ replaced by $b, v_{i}, e_{it}$, respectively. However, unlike $x_{it}$, there are no missing values in $X_{it}$, and this is the most important difference.

{\bf Note 1.} The original GLMM is conditional on the missing data patterns. The same is true for the GLMM ``completed'' for the missing covariates, that is, (\ref{eq:GLMM_pred_2}). In other words, the missing data indicators, such as $1_{(M_{it,2}\neq\emptyset)}$, are considered non-random.

{\bf Note 2.} The proposed mixed-effects modeling, (\ref{eq:mis_decomp}) and (\ref{eq:mc_model}), can be potentially misspecified. The misspecification is taken into consideration in our simulation study in Section 4. In the special case of LMM, the normality assumption about the random effects and errors can be weakened to unspecified distribution (e.g., Jiang and Nguyen 2021, sec. 1.2.2).

This is what we call a random-effects approach. Standard GLMM analysis is then applied to (\ref{eq:GLMM_pred_2}) to estimate the fixed effects, $b$, and variance components involved in the extended GLMM. To see what the extended variance components are, recall $v_{i}=(\alpha_{i}',\gamma_{i})'\sim N(0,H)$ with $H={\rm diag}(G,\sigma_{\gamma}^{2})$, and $e_{it}=\epsilon_{it}+1_{(M_{it,2}\neq\emptyset)}\delta_{it}$. It follows that ${\rm E}(e_{it})=0$ and
$${\rm cov}(e_{is},e_{it})={\rm cov}(\epsilon_{is},\epsilon_{it})+1_{(M_{it,2}\neq\emptyset)}\sigma_{\delta}^{2}1_{(s=t)}.$$
Note that ${\rm cov}(\epsilon_{is},\epsilon_{it})$ is a known function of $\rho$. Thus, the extended variance components are $\psi=(G,\sigma_{\gamma}^{2},\rho,\sigma_{\delta}^{2})$. Typically, the main interest in longitudinal data analysis is $\beta$ and associated mean function; however, the variance components need to be estimated in order to assess uncertainty in the $\beta$ estimator, which is used in inferential analysis. The resulting estimators of $b$ and $\psi$ are called completed-covariates estimators (CCE), denoted by $\hat{b}$ and $\hat{\psi}$, respectively. The name CCE refers to that the estimators are based on the GLMM with completed covariates, (\ref{eq:GLMM_pred_2}). In contrast, the estimators of $\beta$, $G$, etc. based on the complete-data-only-analysis, denoted by $\tilde{\beta}$, $\tilde{G}$, respectively, are called complete-data only estimators (CDOE).

{\bf Note 3.} It is known (e.g., Jiang and Nguyen 2021, sec. 4.1) that maximum likelihood (ML) analysis of GLMM is computationally challenging. However, at least for GLMMs with clustered random effects, such as those for the analysis of longitudinal data, the computation can be handled with the existing software package (see Section 4.2).

As noted in Section 1, an important question is whether there is a gain of CCE over CDOE. Specifically, we are comparing the $\beta$ component of the CCE, $\hat{b}$, denoted by $\hat{\beta}$, and the CDOE, $\tilde{\beta}$, and ask if there is any gain of $\hat{\beta}$ over $\tilde{\beta}$. Results of our simulation studies, presented in Sections 4.1 and 4.2, suggest that the answer is yes. Furthermore, the positive answer can be justified theoretically. To do so, we first establish a theorem regarding the asymptotic normality of the maximum likelihood estimator (MLE) of $\beta$ under the GLMM.

Consider the GLMM introduced in Section 1 but with random $x_{it}$. Suppose that $x_{it}'=(x_{itA}',x_{itB}')$, where $x_{itA}=z_{it}$, and, correspondingly, $\beta=(\beta_{A}',\beta_{B}')'$, such that
\begin{equation}
	x_{it}'\beta=x_{itA}'\beta_{A}+x_{itB}'\beta_{B}=z_{it}'\beta_{A}+x_{itB}'\beta_{B}.\label{eq:fixed_decom}
\end{equation}
Let $R_{itB}$ denote the lower-right block matrix of $[{\rm E}\{b''(\eta_{it})x_{it}x_{it}'|\alpha_{i}\}]^{-1}$ corresponding to $x_{itB}$. Define $\bar{n}=m^{-1}\sum_{i=1}^{m}n_{i}$ (recall $n_{i}=|T_{i}|)$. Also define $a\wedge b=\min(a,b)$, $a\vee b=\max(a,b)$. Let $\lambda_{\min}(M)$ and $\lambda_{\max}(M)$ denote the smallest and largest eigenvalues of matrix $M$, respectively, and $\|M\|=\{\lambda_{\max}(M'M)\}^{1/2}$ be the spectral norm of $M$. The following result can be derived from Theorem 1 of Jiang, Wand, and Bhaskaran (2022) and its proof.

{\bf Theorem 1.} Suppose that (i) the GLMM has canonical link, and $a_{it}=1$; (ii) $m\rightarrow\infty$, $n_{i}\rightarrow\infty$ such that $n_{i}/\bar{n}\rightarrow c_{i}$ for some constant $c_{i}\in(0,\infty)$, $1\leq i\leq m$, and $\bar{n}/m\rightarrow 0$; (iii)
$${\rm E}\left\{\frac{{\rm E}[(1\vee|x_{it}|)^{8}\{1\vee b''(\eta_{it})\}^{2}|\alpha_{i}]}{(1\wedge\lambda_{\min}[{\rm E}\{b''(\eta_{it})x_{itA}x_{itA}'|\alpha_{i}\}])^{2}}\right\}$$
are bounded, where $|v|=(v'v)^{1/2}$ denotes the Euclidean norm of vector $v$, and (iv)
\begin{equation}
	\frac{1}{N}\sum_{i=1}^{m}\sum_{t\in T_{i}}{\rm E}(R_{itB}^{-1})\longrightarrow\Gamma,\label{eq:gamma_lim}
\end{equation}
where $N=\sum_{i=1}^{m}n_{i}$ is the total sample size and $\Gamma$ is a positive definite matrix. Then, we have
\begin{equation}
	\left[\begin{array}{c}\sqrt{m}(\hat{\beta}_{A}-\beta_{A0})\\
		\sqrt{N}(\hat{\beta}_{B}-\beta_{B0})\end{array}\right]\stackrel{\rm d}{\longrightarrow}N\left[\left(\begin{array}{c}0\\
		0\end{array}\right),\left(\begin{array}{cc}G&0\\
		0&\phi\Gamma^{-1}\end{array}\right)\right],\label{eq:a_normal}
\end{equation}
where $G={\rm Var}(\alpha_{i})$, $\hat{\beta}=(\hat{\beta}_{A}',\hat{\beta}_{B}')'$ is the MLE of $\beta$, and $\beta_{0}=(\beta_{A0}',\beta_{B0}')'$ is the true $\beta$.

Theorem 1 suggests that we may compare the efficiency gain of $\hat{\beta}$ over $\tilde{\beta}$ by comparing their asymptotic covariance matrix (ACM) associated with (\ref{eq:a_normal}). Specifically, (\ref{eq:a_normal}) implies that ACM of $\sqrt{N}(\hat{\beta}_{B}-\beta_{B0})$ is $\phi\Gamma^{-1}$; therefore, the ACM of $\hat{\beta}_{B}-\beta_{B0}$ is
\begin{equation}\frac{\phi\Gamma^{-1}}{N}\approx\phi\left\{\sum_{i=1}^{m}\sum_{t\in T_{i}}{\rm E}(R_{itB}^{-1})\right\}^{-1}\equiv\phi\Sigma^{-1},
\end{equation}
by (\ref{eq:gamma_lim}), with $\Sigma$ defined in an obvious way. Thus, the inverse of the ACM (IACM) of $\hat{\beta}_{B}-\beta_{B0}$ is $\Sigma/\phi$. It follows that we can compare the asymptotic efficiency of $\hat{\beta}$ and $\tilde{\beta}$ by looking at their corresponding IACM, which is $\Sigma/\phi$. Whichever has a larger IACM, or equivalently larger $\Sigma$, is relatively more efficient. Here, for two symmetric matrices $M_{1}$, $M_{2}$, $M_{1}$ is larger than $M_{2}$, denoted by $M_{1}>M_{2}$, if $M_{1}-M_{2}$ is positive definite. Similarly, $M_{1}\geq M_{2}$ means that $M_{1}-M_{2}$ is nonnegative definite.

For $\tilde{\beta}$, the CDOE, this is relatively simpler. Define $O_{x,i}=\{t\in T_{i}:M_{it}=\emptyset\}$, and $O_{x}=\{1\leq i\leq m: O_{x,i}\neq\emptyset\}$. Then, the $\Sigma$ corresponding
to $\tilde{\beta}$ is
\begin{equation}
	\Sigma_{0}=\sum_{i\in O_{x}}\sum_{t\in O_{x,i}}{\rm E}(R_{itB}^{-1}).\label{eq:Sigma_0}
\end{equation}

For $\hat{\beta}$, the CCE, we are looking at the $\beta$ component of $\hat{b}$, the CCE of $b$. We need to first sort out the $\beta$ component from $b$. By definition, we can write
\begin{eqnarray*}&&X_{it}'b=X_{1it}'\beta+X_{2it}'\mu=X_{1itA}'\beta_{A}
	+X_{1itB}'\beta_{B}\\
	&&+X_{2itA}'\mu_{A}+X_{2itB}'\mu_{B}=X_{itA}'b_{A}+X_{itB}'b_{B},
\end{eqnarray*}
with $X_{itA}'=(X_{1itA}',X_{2itA}')$, $X_{itB}'=(X_{1itB}',X_{2itB}')$, $b_{A}=(\beta_{A}',\mu_{A}')'$, and $b_{B}=(\beta_{B}',\mu_{B}')'$.
In other words, $X_{itA}$ is the part of $X_{it}$ overlapping with $Z_{it}$. Thus, the $R_{itB}$ corresponding to $b_{B}$, denoted by $R_{itB,b}$, is the lower-right block matrix of $[{\rm E}\{b''(\eta_{it})X_{it}X_{it}'|v_{i},e_{it}\}]^{-1}$ corresponding to $X_{itB}$. Note that ${\rm E}\{b''(\eta_{it})X_{itA}X_{itA}'|v_{i},e_{it}\}=$
\begin{equation}
	\left[\begin{array}{cc}{\rm E}\{b''(\eta_{it})X_{itA}X_{itB}'|v_{i},e_{it}\}&{\rm E}\{b''(\eta_{it})X_{it}X_{it}'|v_{i},e_{it}\}\\
		{\rm E}\{b''(\eta_{it})X_{itB}X_{itA}'|v_{i},e_{it}\}&{\rm E}\{b''(\eta_{it})X_{itB}X_{itB}'v_{i},e_{it}\}\end{array}\right]=\left(\begin{array}{cc}M_{11}&M_{12}\\
		M_{21}&M_{22}\end{array}\right),\label{eq:M_def}
\end{equation}
and, by a block-matrix inversion formula (e.g., Lu and Shiou 2002), we have
$$\left(\begin{array}{cc}M_{11}&M_{12}\\
	M_{21}&M_{22}\end{array}\right)^{-1}=\left[\begin{array}{cc}*&*\\
	*&(M_{22}-M_{21}M_{11}^{-1}M_{12})^{-1}\end{array}\right].$$
It follows that $R_{itB,b}^{-1}=M_{22}-M_{21}M_{11}^{-1}M_{12}\equiv Q_{it}$. Thus, the $\Sigma$ corresponding to the $b_{B}$ is
\begin{equation}
	\Sigma_{B,b}=\sum_{i=1}^{m}\sum_{t\in T_{i}}{\rm E}(R_{itB,b}^{-1})=\sum_{i=1}^{m}\sum_{t\in T_{i}}{\rm E}(Q_{it}),\label{eq:Sigma_Bb}
\end{equation}
where $M_{st}, s,t=1,2$ are defined via (\ref{eq:M_def}). Now write
$$Q_{it}=\left(\begin{array}{cc}Q_{it,11}&Q_{it,12}\\
	Q_{it,21}&Q_{it,22}\end{array}\right),$$
where $Q_{it,11}$ corresponds to the $\beta_{B}$ part of $b_{B}$. It follows that
$$\Sigma_{B,b}=\left[\begin{array}{cc}\sum_{i=1}^{m}\sum_{t\in T_{i}}{\rm E}(Q_{it,11})&\sum_{i=1}^{m}\sum_{t\in T_{i}}{\rm E}(Q_{it,12})\\
	\sum_{i=1}^{m}\sum_{t\in T_{i}}{\rm E}(Q_{it,21})&\sum_{i=1}^{m}\sum_{t\in T_{i}}{\rm E}(Q_{it,22})\end{array}\right]=\left(\begin{array}{cc}S_{11}&S_{12}\\
	S_{21}&S_{22}\end{array}\right),$$
$S_{jk}, j,k=1,2$ defined in obvious ways. Thus, again by the matrix inversion formula (alternative expression; e.g., Lu and Shiou 2002), we have
\begin{equation}
	\Sigma_{B,b}^{-1}=\left[\begin{array}{cc}(S_{11}-S_{12}S_{22}^{-1}S_{21})^{-1}&*\\
		*&*\end{array}\right].\label{eq:Sig_Bb_inv}
\end{equation}

The point is that, by similar arguments as above, $\phi\Sigma_{B,b}^{-1}$ is the ACM of
$$\hat{b}_{B}-b_{B0}=\left(\begin{array}{c}\hat{\beta}_{B}-\beta_{B0}\\
	\hat{\mu}_{B}-\mu_{B0}\end{array}\right).$$
It follows that the upper-left block matrix of $\phi\Sigma_{B,b}^{-1}$, that is, $\phi(S_{11}-S_{12}S_{22}^{-1}S_{21})^{-1}$, is the ACM of $\hat{\beta}_{B}-\beta_{B0}$. Therefore, the IACM of $\hat{\beta}_{B}-\beta_{B0}$ is $(S_{11}-S_{12}S_{22}^{-1}S_{21})/\phi$. It follows that the $\Sigma$ corresponding to $\hat{\beta}$ is
\begin{eqnarray}
	\Sigma_{1}&=&S_{11}-S_{12}S_{22}^{-1}S_{21}\nonumber\\
	&=&\sum_{i=1}^{m}\sum_{t\in T_{i}}{\rm E}(Q_{it,11})\nonumber\\
	&&-\left\{\sum_{i=1}^{m}\sum_{t\in T_{i}}{\rm E}(Q_{it,12})\right\}
	\left\{\sum_{i=1}^{m}\sum_{t\in T_{i}}{\rm E}(Q_{it,22})\right\}^{-1}
	\left\{\sum_{t\in T_{i}}{\rm E}(Q_{it,21})\right\}.\label{eq:Sigma_1}
\end{eqnarray}

After the above preparation, we are almost ready to state and prove our main theoretical result. Recall the notation $R_{itB}$ introduced below (\ref{eq:fixed_decom}). The notation depends on the $x$, $z$, and the random variable $\alpha_{i}$ in the conditioning. To highlight this, we write $R_{itB}=R_{it}(x,z,\alpha_{i})$. Here, we have an extra random variable, $\epsilon_{it}$, in $\eta_{it}$ [see (\ref{eq:GLMM_pred_1})]. Thus, the extended notation is $R_{itB}(x,z,\alpha_{i},\epsilon_{it})$. We assume the following additional regularity conditions:
\begin{eqnarray}
	&&\limsup\left\{\sum_{i=1}^{m}\sum_{t\in T_{i}}{\rm E}(Q_{it,11})\right\}^{-1/2}\left\{\sum_{i\in O_{x}}\sum_{t\in O_{x,i}}{\rm E}(R_{itB}^{-1})\right\}\nonumber\\
	&&\times\left\{\sum_{i=1}^{m}\sum_{t\in T_{i}}{\rm E}(Q_{it,11})\right\}^{-1/2}<I,\label{eq:thm2_3}
\end{eqnarray}
where $R_{itB}=R_{itB}(x,z,\alpha_{i},e_{it})$, $Q_{it,11}$ is defined above, and $I$ is the identity matrix of the same dimension as the left side of (\ref{eq:thm2_3}); and
\begin{eqnarray}
	&&\left\{\sum_{i=1}^{m}\sum_{t\in T_{i}}{\rm E}(Q_{it,11})\right\}^{-1/2}
	\left\{\sum_{i=1}^{m}\sum_{t\in T_{i}}{\rm E}(Q_{it,12})\right\}
	\left\{\sum_{i=1}^{m}\sum_{t\in T_{i}}{\rm E}(Q_{it,22})\right\}^{-1}
	\nonumber\\
	&&\times\left\{\sum_{i=1}^{m}\sum_{t\in T_{i}}{\rm E}(Q_{it,21})\right\}\left\{\sum_{i=1}^{m}\sum_{t\in T_{i}}{\rm E}(Q_{it,11})\right\}^{-1/2}=o(1).\label{eq:thm2_4}
\end{eqnarray}
To see these conditions are reasonable, first note that the summation range $\sum_{i\in O_{x}}\sum_{t\in O_{x,i}}$ is a subset of that of $\sum_{i=1}^{m}\sum_{t\in T_{i}}$; in fact, the former could be a much smaller subset of the latter if there are many missing covariates. On the other hand, ${\rm E}(R_{itB}^{-1})$ and ${\rm E}(Q_{it,11})$ are the same kind of quantities [see (\ref{eq:Sigma_Bb}) and the block-matrix expression of $Q_{it}$ below it]. Therefore, one expects $\sum_{i\in O_{x}}\sum_{t\in O_{x,i}}{\rm E}(R_{itB}^{-1})$ to be (much) smaller than $\sum_{i=1}^{m}\sum_{t\in T_{i}}{\rm E}(Q_{it,11})$, hence (\ref{eq:thm2_3}) to hold. On the other hand, by the definition of $X_{it}$, one expects many zero terms in the sum, $\sum_{i=1}^{m}\sum_{t\in T_{i}}{\rm E}(Q_{it,12})$, but not as many zero terms in the sum, $\sum_{i=1}^{m}\sum_{t\in T_{i}}{\rm E}(Q_{it,22})$. Also note that, by the definitions, $\sum_{i=1}^{m}\sum_{t\in T_{i}}{\rm E}(Q_{it,11})$ and $\sum_{i=1}^{m}\sum_{t\in T_{i}}{\rm E}(Q_{it,22})$ are both positive definite. The implication is that $\sum_{i=1}^{m}\sum_{t\in T_{i}}{\rm E}(Q_{it,12})=\{\sum_{i=1}^{m}\sum_{t\in T_{i}}{\rm E}(Q_{it,21})\}'$ is at most of the same order as $\sum_{i=1}^{m}\sum_{t\in T_{i}}{\rm E}(Q_{it,22})$ but is of lower order than $\sum_{i=1}^{m}\sum_{t\in T_{i}}{\rm E}(Q_{it,11})$; therefore, (\ref{eq:thm2_4}) is expected to hold.

{\bf Theorem 2.} Suppose that conditions (i), (ii) of Theorem 1 hold; condition (iii) of Theorem 1 holds with $\alpha_{i}$ in the conditioning replaced by $\alpha_{i},\epsilon_{it}$, and it holds with $x,z$ replaced by $X,Z$ and $\alpha_{i}$ replaced by $v_{i}, e_{it}$. Also, suppose
\begin{eqnarray}
	\frac{1}{N_{\rm o}}\sum_{i\in O_{x}}\sum_{t\in O_{x,i}}{\rm E}\{R_{itB}(x,z,\alpha_{i},\epsilon_{it})^{-1}\}&\longrightarrow&\Gamma_{0},\label{eq:thm2_1}\\
	\frac{1}{N}\sum_{i=1}^{m}\sum_{t\in T_{i}}{\rm E}\{R_{itB}(X,Z,v_{i},e_{it})^{-1}\}&\longrightarrow&\Gamma_{1},
\end{eqnarray}
where $\Gamma_{0},\Gamma_{1}$ are positive definite, and (\ref{eq:thm2_3}), (\ref{eq:thm2_4}) hold. Then, we have
\begin{equation}
	\Sigma_{1}>\Sigma_{0}\label{eq:thm2_5}
\end{equation}
for large $m$ and $n_{i}$'s, where $\Sigma_{0}$ is the IACM of $\tilde{\beta}$, the CDOE, and $\Sigma_{1}$ is the IACM of $\hat{\beta}$, the CCE. In other words, the CCE is asymptotically more efficient than the CDOE.

{\bf Proof.} By extending the proof of Theorem 1 (see Jiang {\it et al.} 2022), it can be shown that the asymptotic normality result, (\ref{eq:a_normal}), holds, for both the CDOE and the CCE. Then, following the above derivations, the IACMs of $\tilde{\beta}$ and $\hat{\beta}$ are given by (\ref{eq:Sigma_0}), with $R_{itB}=R_{itB}(x,z,\alpha_{i},\epsilon_{it})$, and (\ref{eq:Sigma_1}), respectively. Furthermore, by (\ref{eq:thm2_3}), there is $\delta>0$ such that, for large $m$ and $n_{i}$'s, we have
$$\left\{\sum_{i=1}^{m}\sum_{t\in T_{i}}{\rm E}(Q_{it,11})\right\}^{-1/2}\Sigma_{0}\left\{\sum_{i=1}^{m}\sum_{t\in T_{i}}{\rm E}(Q_{it,11})\right\}^{-1/2}<(1-\delta)I.$$
It follows, by (\ref{eq:thm2_4}) and (\ref{eq:thm2_3}), that, for large $m$ and $n_{i}$'s, we have
\begin{eqnarray*}
	&&\left\{\sum_{i=1}^{m}\sum_{t\in T_{i}}{\rm E}(Q_{it,11})\right\}^{-1/2}\Sigma_{1}\left\{\sum_{i=1}^{m}\sum_{t\in T_{i}}{\rm E}(Q_{it,11})\right\}^{-1/2}\\
	&=&I-\left\{\sum_{i=1}^{m}\sum_{t\in T_{i}}{\rm E}(Q_{it,11})\right\}^{-1/2}
	\left\{\sum_{i=1}^{m}\sum_{t\in T_{i}}{\rm E}(Q_{it,12})\right\}
	\left\{\sum_{i=1}^{m}\sum_{t\in T_{i}}{\rm E}(Q_{it,22})\right\}^{-1}
	\nonumber\\
	&&\times\left\{\sum_{i=1}^{m}\sum_{t\in T_{i}}{\rm E}(Q_{it,21})\right\}\left\{\sum_{i=1}^{m}\sum_{t\in T_{i}}{\rm E}(Q_{it,11})\right\}^{-1/2}\\
	&=&(1-\delta)I+\delta I+o(1)\\
	&>&\left\{\sum_{i=1}^{m}\sum_{t\in T_{i}}{\rm E}(Q_{it,11})\right\}^{-1/2}\Sigma_{0}\left\{\sum_{i=1}^{m}\sum_{t\in T_{i}}{\rm E}(Q_{it,11})\right\}^{-1/2}+\delta I+o(1)\\
	&\geq&\left\{\sum_{i=1}^{m}\sum_{t\in T_{i}}{\rm E}(Q_{it,11})\right\}^{-1/2}\Sigma_{0}\left\{\sum_{i=1}^{m}\sum_{t\in T_{i}}{\rm E}(Q_{it,11})\right\}^{-1/2}+\frac{\delta}{2}I.
\end{eqnarray*}
In other words, we have
$$\left\{\sum_{i=1}^{m}\sum_{t\in T_{i}}{\rm E}(Q_{it,11})\right\}^{-1/2}(\Sigma_{1}-\Sigma_{0})\left\{\sum_{i=1}^{m}\sum_{t\in T_{i}}{\rm E}(Q_{it,11})\right\}^{-1/2}\geq\frac{\delta}{2}I,$$
hence, for for large $m$ and $n_{i}$'s, we have [e.g., Jiang 2022, \S{5.3.1} (ii)]
$$\Sigma_{1}-\Sigma_{0}\geq\frac{\delta}{2}\sum_{i=1}^{m}\sum_{t\in T_{i}}{\rm E}(Q_{it,11})>0.$$
This completes the proof.
\section{Missing responses and covariates}
\hspace{4mm}
We now consider the case where there are missing values in both the covariates and the responses. We are going to show that (I) the cases with missing covariates but observed responses help, and that (II) the cases with missing responses do not help. (I) is not surprising because this case has been established in the previous section, if one only considers the cases with observed responses, regardless of the missing status of the covariates. To demonstrate (II), we are going to consider a special case, that is, the case of LMM, and show that the cases with missing responses do not help even in this case. The point is that a method that works should work in a simple, ideal situation (otherwise, forget about the method).

In the cases of LMM, it is more convenient to use the matrix expression,
\begin{equation}
	y=X\beta+Z\alpha+\epsilon,\label{eq:LMM_eq}
\end{equation}
where $y=(y_{i})_{1\leq i\leq m}$ with $y_{i}=(y_{it})_{t\in T_{i}}$, $X=(X_{i})_{1\leq i\leq m}$ with $X_{i}=(x_{it}')_{t\in T_{i}}$, $Z={\rm diag}(Z_{i}$, $1\leq i\leq m)$ with $Z_{i}=(z_{it}')_{t\in T_{i}}$, and $\epsilon=(\epsilon_{i})_{1\leq i\leq m}$ with $\epsilon_{i}=(\epsilon_{it})_{t\in T_{i}}$. Also assume $\epsilon\sim N(0,\sigma_{\epsilon}^{2}I_{N})$ with $N=\sum_{i=1}^{m}n_{i}$ for simplicity.
First note that the random-effects approach of handling the missing covariates described in the previous section, works only on the right side of the equation (\ref{eq:GLMM_pred_1})---it has nothing to do with the $y$, whether it is observed or not. Thus, as a first step, we apply the random-effects procedure of the previous section to ``complete'' the missing covariates; as a result, we can assume that (\ref{eq:GLMM_pred_2}) holds and there are no missing values in the covariates, which can be written in the matrix form as
\begin{equation}
	y=X_{1}\beta+X_{2}\mu+Zv+e=Xb+Zv+e.\label{eq:LMM_mat}
\end{equation}

Now let $y_{\rm o}$ and $y_{\rm m}$ denote the sub-vectors of observed and missing responses, respectively. Without loss of generality, let $y=(y_{\rm o}',y_{\rm m}')'$. Let $X_{\rm o}$, $Z_{\rm o}$ and $e_{\rm o}$ be the sub-matrices and sub-vector of $X$, $Z$ and $e$ corresponding to $y_{\rm o}$, and $X_{\rm m}$, $Z_{\rm m}$ and $e_{\rm m}$ those corresponding to $y_{\rm m}$, respectively. For simplicity, assume the covariates are fixed, that is, the LMM is conditional on the covariates. If the data are normally distributed, it can be shown that (e.g., Jiang and Nguyen 2021, p. 325) the best predictor (BP) of $y_{\rm m}$ based on $y_{\rm o}$, in the sense of minimizing the mean squared prediction error (MSPE), is the conditional expectation of $y_{\rm m}$ given $y_{\rm o}$, which is
\begin{equation}
	\tilde{y}_{\rm m}=X_{\rm m}b+C_{\rm m,o}V_{\rm o}^{-1}(y_{\rm o}-X_{\rm o}b),\label{eq:BP_y_m}
\end{equation}
where $C_{\rm m,o}={\rm Cov}(y_{\rm m},y_{\rm o})=Z_{\rm m}{\rm Var}(v)Z_{\rm o}'$ and $V_{\rm o}={\rm Var}(y_{\rm o})={\rm Var}(e_{\rm o})+Z_{\rm o}{\rm Var}(v)Z_{\rm o}'$. Let $D_{i}={\rm diag}\{1_{(M_{it,2}\neq\emptyset)},t\in T_{i}\}$, and $D={\rm diag}(D_{i},1\leq i\leq m)$. It is easy to see that $e=\epsilon+D\delta$, where $\delta=(\delta_{i})_{1\leq i\leq m}$ with $\delta_{i}=(\delta_{it})_{t\in T_{i}}$. It follows that ${\rm Var}(e)=\sigma_{\epsilon}^{2}I_{N}+\sigma_{\delta}^{2}D$. Thus, we have ${\rm Var}(e_{\rm o})=\sigma_{\epsilon}^{2}I_{N_{\rm o}}+\sigma_{\delta}^{2}D_{\rm o}$ with $N_{\rm o}$ being the number of rows of $y_{\rm o}$ and $D_{\rm o}$ being the submatrix of $D$ whose rows correspond to $y_{\rm o}$. Furthermore, it can be shown that ${\rm Var}(v)=I_{m}\otimes H$, where $\otimes$ denote the Kronecker product and $H={\rm daig}(G,\sigma_{\gamma}^{2})$. It follows that we have
\begin{eqnarray}
	C_{\rm m,o}&=&Z_{\rm m}(I_{m}\otimes H)Z_{\rm o}',\label{eq:BP_y_m1}\\
	V_{\rm o}&=&\sigma_{\epsilon}^{2}I_{N_{\rm o}}+\sigma_{\delta}^{2}D_{\rm o}+Z_{\rm o}(I_{m}\otimes H)Z_{\rm o}'.\label{eq:BP_y_m2}
\end{eqnarray}
With (\ref{eq:BP_y_m})--(\ref{eq:BP_y_m2}), the BP can be computed, if the parameters involved are known. In practice, the unknown parameters involved in the BP, namely, the $b$ and $\psi$ in (\ref{eq:LMM_mat}) are replaced by their estimators. Naturally, those parameters are estimated by fitting the sub-model,
\begin{equation}
	y_{\rm o}=X_{\rm o}b+Z_{\rm o}v+e_{\rm o}.\label{eq:sub_LMM_o}
\end{equation}
When the parameters involved in (\ref{eq:BP_y_m})--(\ref{eq:BP_y_m2})
are replaced by their estimators based on (\ref{eq:sub_LMM_o}), denoted by $\hat{b}_{\rm o}$ and $\hat{\psi}_{\rm o}$, the result is called the empirical BP (EBP) of $y_{\rm m}$, denoted by $\hat{y}_{\rm m}$. This is called a random-effects predictive (Rep) approach for an obvious reason.

We now go back to fitting (\ref{eq:LMM_mat}), with $y_{\rm m}$ replaced by $\hat{y}_{\rm m}$; in other words, with $y$ replaced by $\tilde{y}=(y_{\rm o}',\hat{y}_{\rm m}')'$. This leads to what we call completed-covariates predictive estimator (CCPE).

Note that there are two differences between CCPE and CDOE. First, some covariates corresponding to $y_{\rm o}$ may be missing; the records corresponding to the missing covariates of $y_{\rm o}$ cannot be used in CDOE. Second, in addition to $y_{\rm o}$ and $X_{\rm o}$, we also have $X_{\rm m}$, which may provide further information. The difference between CCPE and CCE is more easily seen: CCE is obtained by simply fitting (\ref{eq:sub_LMM_o}), while CCPE is obtained by fitting (\ref{eq:LMM_mat}) with $y$ replaced by $\tilde{y}$.

Again, a key question is whether one gains anything by doing CCPE over CDOE, or CCE. In the simpler case of linear regression, which corresponds to (\ref{eq:LMM_eq}) without the term $Z\alpha$, and assuming there is no missing value in $X_{\rm o}$, it can be shown that CCPE and CCE are exactly the same; thus, in this special case, there is no loss or gain of CCPE over CCE. However, this could be due to the fact that $y_{\rm o}$ and $y_{\rm m}$ are independent in this case. Under a LMM, there are correlations between $y_{\rm o}$ and $y_{\rm m}$. Will this make a difference?

Our simulation study, presented in Section 4.3, suggests otherwise, that is, there is no gain of CCPE over CCE. This no-gain can also be explained theoretically, as we do next.

{\bf Theoretical explanation:} Let us first consider a simpler case, in which the variance components involved in $V_{0}$ and $C_{m,{\rm o}}$ are known [see (\ref{eq:BP_y_m})]. write $\zeta_{\rm o}=Z_{\rm o}v+e_{\rm o}$ [see (\ref{eq:sub_LMM_o})]. The EBP of $y_{\rm m}$, which is now the best linear unbiased predictor (BLUP), is (\ref{eq:BP_y_m}) with $b$ replaced by its MLE under model (\ref{eq:sub_LMM_o}), that is, $\tilde{b}=(X_{\rm o}'V_{\rm o}^{-1}X_{\rm o})^{-1}X_{\rm o}'V_{\rm o}^{-1}y_{\rm o}$. Then, with the Rep approach, we have
\begin{equation}\tilde{y}=\left(\begin{array}{c}y_{\rm o}\\
		\tilde{y}_{\rm m}\end{array}\right)=\left(\begin{array}{c}I_{N_{\rm o}}\\
		Q\end{array}\right)y_{\rm o},\label{eq:y_tilde}
\end{equation}
where $Q=\{C_{\rm m,o}+(X_{\rm m}-C_{\rm m,o}V_{\rm o}^{-1}X_{\rm o})(X_{\rm o}'V_{\rm o}^{-1}X_{\rm o})^{-1}X_{\rm o}'\}V_{\rm o}^{-1}$ and $N_{\rm o}={\rm dim}(y_{\rm o})$. It is known (e.g., Jiang 1998) that, under regularity conditions, the asymptotic covariance matrix of the MLE of $\beta$ can be expressed in the form of $(X'V^{-1}X)^{-1}$, where $X$ is the matrix of covariates under the LMM and $V$ is the covariance matrix of $y$ under the LMM. However, it can be shown that the covariance matrix of $\tilde{y}$ is singular. In fact, we have
$$\tilde{V}\equiv{\rm Var}(\tilde{y})=\left(\begin{array}{cc}V_{\rm o}&V_{\rm o}\\
	C_{\rm m,o}&C_{\rm m,o}\end{array}\right)\left(\begin{array}{cc}I_{N_{\rm o}}&0\\
	0&V_{\rm o}^{-1}C_{\rm m,o}'\end{array}\right).$$
Thus, although one is fitting the LMM (\ref{eq:LMM_mat}), one cannot use $(X'\tilde{V}^{-1}X)^{-1}$ to approximate the asymptotic covariance matrix of the resulting estimator of $b$, whose $\beta$ component is the CCPE.

In order to still make a comparison, note that, when the variance components are known, the MLE of $\beta$ is the same as the best linear unbiased estimator (BLUE). Thus, under this situation, the CCPE of $\beta$ is a linear function of $\tilde{y}$. It then follows, by (\ref{eq:y_tilde}), that the CCPE is a linear function of $y_{\rm o}$. This estimator cannot be better than the BLUE of $\beta$ based on $y_{\rm o}$ (by the definition of BLUE), which is the CCE of $\beta$.

As for the case where the variance components are unknown, they are estimated, say, by the ML estimator. As noted earlier, there may be misspecification of the proposed mixed-effects modeling, (\ref{eq:mis_decomp}) and (\ref{eq:mc_model}) (see Note 2 in Section 2). Nevertheless, following the arguments of White (1982), it can be shown that, under the misspecified LMM, the MLE of the variance components still converge (in probability) to ``something''. It can then be shown that, under regularity conditions, the CCPE is approximately a linear function of $y_{\rm o}$, and the CCE is approximately the BLUE of $\beta$ based on $y_{\rm o}$. It then follows that, asymptotically, CCPE cannot be a superior estimator than CCE.

What is more, there is a potential danger when using CCPE. Namely, using the CCPE approach, the ``sample size'' may be much larger. While we know this does not help, a software package does not (or may not), which may report a false efficiency gain as well as statistical significance. This is illustrated using a real-data example in Section 5.1.
\section{Simulation studies}
\hspace{4mm}
We carry out a series of simulation studies to demonstrate the efficiency gain of CCE and the no (further) efficiency gain of CCPE, as well as their comparison with MICE.
\subsection{Efficiency gain of CCE under LMM}
\hspace{4mm}
We begin with a scenario considered in Section 2, in which there are missing covariates but no missing responses. The complete data are generated under the following LMM:
\begin{equation}
	y_{it}=\beta_{0}+\beta_{1}x_{1,i}+\beta_{2}x_{2,i}+\beta_{3}t+\beta_{4}x_{4,it}+\alpha_{i}+\epsilon_{it},\label{sim_model1}
\end{equation}
$i=1,\dots,m$, $t=1,\dots,5$, where $\beta_{0}=1.0$, $\beta_{1}=0.5$, and $\beta_{2}=\beta_{3}=\beta_{4}=0.2$. Here, $x_{1,i}$ is generated independently from Uniform$[0,1]$, $x_{2,i}$ is an indicator variable generated independently from Bernoulli$(0.5)$, $x_{4,it}$ is generated independently from $N(1,1)$, and $x_{1}$, $x_{2}$, $x_{4}$ are independent. Furthermore, the random effects $\alpha_{i}$ are generated independently from $N(0,0.3^2)$; the errors $\epsilon_{it}$ are generated independently from $N(0,0.1^2)$, and $x$, $\alpha$, $\epsilon$ are independent.

Here, $x_{2}$ corresponds to a ``treatment indicator'' (1 for treatment; 0 for control); there is another covariate, $x_{3,t}=t$, corresponding to $\beta_{3}$, with $t$ being the time. These two covariates are known by design, so no missing values are involved.

Furthermore, $x_{1}$ is a subject-level covariate corresponding to a baseline percentage, and $x_{4}$ is a time-varying covariate; those two covariates have missing values. Specifically, let $I_{1,i}$ and $I_{4,it}$ denote the missing value indicators for $x_{1,i}$ and $x_{4,it}$, respectively. To incorporate the missing at random mechanism, $I_{1,i}$ are generated independently from the Bernoulli$(p_i)$ distribution, where $p_i={\rm logit}(3 x_{2,i}-2)$; $I_{4,it}$ are generated independently from the Bernoulli$(qt)$ distribution, $t=1,\dots,5$; and $I_{1}$ and $I_{4}$ are independent. For the values of $q$, we set $q=0.15$. We consider $m=40$, $m=100$, and $m=400$ in our simulation study.

We compare the performance of CCE with that of CDOE in estimating $\beta=(\beta_{k})_{0\leq k\leq 4}$, which is a subvector of $b$. Note that $\beta$ is a common vector of fixed effects in both (\ref{eq:GLMM_pred_1}) and (\ref{eq:GLMM_pred_2}). Also among the comparing methods is the estimator of $\beta$ based on the MICE imputation, denoted by MICE (see Section 1). MICE was performed with m = 50 imputations using the \textit{mice} package in R, employing predictive mean matching imputation methods along with the Namard-Rubin pooling rule. We also provide the estimator of $\beta$ with the Full dataset without any missing as a reference, denoted by FULL. We compare the simulated mean, standard deviation (SD), and mean squared error (MSE) of different estimators based on $N_{\rm sim}=1000$ simulation runs.

\begin{table}[htp]
	\centering
	\resizebox{0.9\textwidth}{!}{
		\begin{tabular}{c|c|ccc|ccc|ccc}
			& &\multicolumn{3}{c|}{$m=40$} &\multicolumn{3}{c|}{$m=100$} &\multicolumn{3}{c}{$m=400$}\\
			Parameter & Methods& MSE($10^{-3}$) & Mean & SD & MSE($10^{-3}$) & Mean & SD & MSE($10^{-3}$) & Mean & SD \\
			\hline
			$\beta_{0}$ & FULL & 24.374& 0.971  & 0.154 & 5.775 & 1.003 & 0.076 & 2.165 & 0.998 & 0.046\\
			& CDOE  & 29.938& 0.985  & 0.173 & 9.621 & 1.001 & 0.098 & 3.111 & 0.998 & 0.056\\ 
			& CCE & 28.596& 0.986 &  0.169 & 8.728 & 1.001 & 0.093 & 2.940 & 0.997 & 0.054\\
			& MICE & 47.136& 1.148  &  0.159 & 39.88 & 1.183 & 0.080 & 3.673 & 1.185 & 0.050\\
			$\beta_{1}$ & FULL & 41.284& 0.509  &  0.204 & 11.603 & 0.492 & 0.108 & 3.035 & 0.501 & 0.055\\
			& CDOE  & 76.363& 0.487  & 0.277 & 22.667 & 0.496 & 0.151 & 5.881 & 0.500 & 0.077\\ 
			& CCE & 70.335& 0.481  &  0.265 & 21.471 & 0.496 & 0.146 & 5.638 & 0.501 & 0.075\\
			& MICE & 78.244& 0.271 &  0.162 & 68.958 & 0.254 & 0.094 & 64.885 & 0.249 & 0.043\\
			$\beta_{2}$ & FULL & 10.444& 0.214  &  0.102 & 3.992 & 0.197 & 0.063 & 0.951 & 0.200 & 0.031\\
			& CDOE  & 33.884&  0.227 & 0.183 & 11.315 & 0.195 & 0.106 & 2.624 & 0.200 & 0.051\\ 
			& CCE & 21.106& 0.231 & 0.142 & 7.256 & 0.195 &  0.085 & 1.739 & 0.200 & 0.042\\
			& MICE & 12.424& 0.214 & 0.111 & 4.803 & 0.195 & 0.069 & 1.169 & 0.200 & 0.034\\\
			$\beta_{3}$ & FULL & 0.428& 0.197  & 0.021 & 0.050 & 0.200 & 0.007 & 0.042 & 0.200 & 0.007\\
			& CDOE  & 0.735& 0.198  &  0.027 & 0.154 & 0.200 & 0.012 & 0.068 & 0.200 & 0.008\\ 
			& CCE & 0.619& 0.197 &  0.025 & 0.116 & 0.200 & 0.010 & 0.058 & 0.200 & 0.007\\
			& MICE & 0.544& 0.196 & 0.023 & 0.117 & 0.199 & 0.010 & 0.059 & 0.200 & 0.007\\
			$\beta_{4}$ & FULL & 0.470&  0.196 &  0.022 & 0.066 & 0.200 & 0.008 & 0.047 & 0.200 & 0.007\\
			& CDOE  &0.868& 0.193  &  0.029 & 0.211 & 0.200 & 0.015 & 0.081 & 0.200 & 0.009\\ 
			& CCE & 0.687& 0.196 & 0.026 & 0.137 & 0.200 & 0.012 & 0.065 & 0.200 & 0.008\\
			& MICE & 4.505& 0.137 &  0.024 & 3.839 & 0.139 & 0.013 & 3.837 & 0.139 & 0.008\\
		\end{tabular}
	}
	\vspace{2mm}
	\caption{Simulated Mean, Standard Deviation (SD), and Mean Squared Error (MSE) of Different Approaches, Parameters and Sample Sizes: Linear Mixed Model with Missing Covariates}
	\label{sim_Res_covariate}
\end{table}

Table \ref{sim_Res_covariate} reports results of parameter estimation for different estimators. It can be seen that the FULL estimator has the lowest MSE across all sample size settings for all $\beta$, which, of course, is reasonable. Among the three estimators handling missing data, MICE is most biased across all sample sizes for the intercept, $\beta_1$, and $\beta_4$, which are the coefficients of covariates with missing values. In contrast, CDOE and CCE produce unbiased estimators for all model parameters. More specifically, CCE demonstrates smaller MSE and smaller SD compared to CDOE. When comparing MICE to CDOE and CCE in estimating coefficients of covariates without missing values, which are $\beta_2$, and $\beta_3$, the MSE of MICE estimators are the smallest. This is because MICE has the lowest estimation variability due to the estimators' smaller SD. In fact, the simulated SD of the MICE estimator is even smaller than that of the FULL estimator for estimating $\beta_1$.

In summary, the simulation results show that CCE performs uniformly better than CDOE across different parameters and sample sizes. The performance of MICE is unstable across different parameters in that it outperforms CCE and CDOE for parameters associated with covariates that do not have missing values, but performs much poorer compared to CCE and CDOE for parameters associated with covariates that have missing values.
\subsection{Efficiency of CCE under mixed logistic model}
\hspace{4mm}
In view of a real-data application to be considered in the next section, we consider a scenario of logistic regression, which does not involve any random effects or additional errors. Similar to the previous subsection, we consider a case in which there are missing covariates but no missing responses. The complete data are generated under the following model:
\begin{equation}
	\begin{split}
		y_{it} &\sim{\rm Bernoulli}(p_{it});\\
		{\rm logit}(p_{it}) & = \eta_{it};\\
		\eta_{it} &= \beta_{0}+\beta_{1}x_{1,i}+\beta_{2}x_{2,i}+\beta_{3}t+\beta_{4}x_{4,it},
		\label{sim_model1_logistic}
	\end{split} 
\end{equation}
where the covariates and the true $\beta$ values are the same as in \eqref{sim_model1}, except that $\beta_0=-1$. The missing values are generated in the same way as in the previous subsection. Again, we consider $m=40$, $m=100$, and $m=400$. The results are presented in Table \ref{sim_Res_covariate_logistic_no_dispersion}.

\begin{table}[htp]
	\centering
	\resizebox{0.9\textwidth}{!}{
		\begin{tabular}{c|c|ccc|ccc|ccc}
			& &\multicolumn{3}{c|}{$m=40$} &\multicolumn{3}{c|}{$m=100$} &\multicolumn{3}{c}{$m=400$}\\
			Parameter & Methods& MSE & Mean & SD & MSE & Mean & SD & MSE & Mean & SD \\
			\hline
			$\beta_{0}$ & FULL & 0.248& -1.029  & 0.497 & 0.100 & -1.007 & 0.316 & 0.024 & -1.002 & 0.156\\
			& CDOE  & 0.854& -1.084  & 0.921 & 0.297 & -1.032 & 0.544 & 0.061 & -1.000 & 0.248\\ 
			& CCE & 0.332& -0.998 &  0.647 & 0.151 & -1.014 & 0.413 & 0.036 & -1.001 & 0.190\\
			& MICE & 0.420& -1.033 &  0.577 & 0.170 & -0.977 & 0.389 & 0.038 & -0.992 & 0.195\\
			$\beta_{1}$ & FULL & 0.287& 0.511  &  0.536 & 0.096 & 0.490 & 0.310 & 0.026 & 0.499 & 0.163\\
			& CDOE  & 1.238& 0.496  & 1.113 & 0.370 & 0.464 & 0.608 & 0.083 & 0.499 & 0.289\\ 
			& CCE & 0.573& 0.495  &  0.757 & 0.193 & 0.476 & 0.439 & 0.046 & 0.497 & 0.216\\
			& MICE & 0.589& 0.462 &  0.767 & 0.212 & 0.453 & 0.459 & 0.062 & 0.486 & 0.249\\
			$\beta_{2}$ & FULL & 0.093& 0.198  &  0.306 & 0.032 & 0.206 & 0.180 & 0.009 & 0.200 & 0.094\\
			& CDOE  & 0.857&  0.045 & 2.416 & 0.174 & 0.204 & 0.417 & 0.038 & 0.198 & 0.195\\ 
			& CCE & 0.184& 0.202 & 0.430 & 0.057 & 0.206 &  0.239 & 0.014 & 0.198 & 0.119\\
			& MICE & 0.111& 0.199 & 0.333 & 0.034 & 0.208 & 0.184 & 0.010 & 0.201 & 0.099\\\
			$\beta_{3}$ & FULL & 0.012& 0.208  & 0.109 & 0.005 & 0.203 & 0.068 & 0.001 & 0.202 & 0.033\\
			& CDOE  & 0.055& 0.228  &  0.233 & 0.018 & 0.212 & 0.133 & 0.004 & 0.203 & 0.062\\ 
			& CCE & 0.027& 0.211 &  0.165 & 0.009 & 0.205 & 0.096 & 0.002 & 0.202 & 0.047\\
			& MICE & 0.012& 0.209 & 0.110 & 0.004 & 0.204 & 0.068 & 0.001 & 0.202 & 0.034\\
			$\beta_{4}$ & FULL & 0.022&  0.203 &  0.150 & 0.009 & 0.201 & 0.095 & 0.002 & 0.198 & 0.046\\
			& CDOE  &0.090& 0.221  &  0.300 & 0.033 & 0.214 & 0.180 & 0.006 & 0.199 & 0.080\\ 
			& CCE & 0.043& 0.207 & 0.209 & 0.017 & 0.205 & 0.132 & 0.003 & 0.197 & 0.060\\
			& MICE & 0.044& 0.193 &  0.217 & 0.019 & 0.191& 0.137 & 0.005 & 0.193 & 0.072\\
		\end{tabular}
	}
	\vspace{2mm}
	\caption{Simulated Mean, Standard Deviation (SD), and Mean Squared Error (MSE) of Different Approaches, Parameters and Sample Sizes: Logistic Regression with Missing Covariates}
	\label{sim_Res_covariate_logistic_no_dispersion}
\end{table}

The overall picture of the results is similar to what was observed in Table \ref{sim_Res_covariate}, although MICE performs relatively much better than in the case of LMM. Specifically, the MSE of FULL is almost always the smallest, which is not surprising. All four methods produce nearly unbiased estimations. In terms of the MSE, among the three methods that need to handle the missing values, CCE has the smallest MSE for estimating $\beta_{0}$, $\beta_{1}$ and $\beta_{4}$; note that $\beta_{1}$ and $\beta_{4}$ correspond to the covariates with missing values. On the other hand, MICE has the smallest MSE for estimating $\beta_{2}$ and $\beta_{3}$, which correspond to the covariates without missing values. It is also seen that CCE significantly outperforms CDOE in terms of the MSE in all cases. In contrast, the MSE results of CCE and MICE are comparable in all cases, regardless of who is performing better.
\subsection{No (further) efficiency gain of CCPE under LMM}
\hspace{4mm}
Another simulation study is carried out to assess the performance of different estimators proposed in Section 3, where both missing covariates and missing responses are present. The complete data are generated under the same LMM as \eqref{sim_model1}. The difference is that now some responses are missing at random. To be more specific, let $I_{y,it}$ denote the missing value indicators for $y_{it}$, which are generated independently from the Bernoulli$(r_{it})$ distribution with
$r_{it}={\rm logit}(3 x_{2,i}+2 x_{4,it}-3)$.
The missing data mechanism of the covariates is the same as in section 4.1. To avoid a relatively high probability of failing to produce the estimates under a smaller sample size, here we only consider $m=100$ and $m=400$ in our simulation study.

We compare the performance of CCPE and CCE with that of CDOE in estimating $\beta$. We also include FULL and MICE in our comparison. In Table \ref{sim_Res_response}, we report the simulated mean, SD, and MSE of estimators based on $N_{sim}=1000$ simulation runs.
\begin{table}[htp]
	\centering
	\resizebox{0.9\textwidth}{!}{
		\begin{tabular}{c|c|ccc|ccc}
			& &\multicolumn{3}{c|}{$m=100$} &\multicolumn{3}{c}{$m=400$}\\
			Parameter & Methods & MSE($10^{-3}$) & Mean & SD & MSE($10^{-3}$) & Mean & SD \\
			\hline
			$\beta_{0}$ & FULL  & 5.733 & 0.997 & 0.076 & 2.188 & 0.999 & 0.047\\
			& CDOE  & 16.296 & 0.995 & 0.128 & 4.413 & 0.998 & 0.066\\ 
			& CCE  & 10.615 & 0.994 & 0.103 & 3.321 & 0.999 & 0.058\\
			& CCPE  & 10.734 & 0.994 & 0.103 & 3.324 & 0.999 & 0.058\\
			& MICE  & 13.935 & 1.034 & 0.113 & 4.814 & 1.030 & 0.063\\
			$\beta_{1}$ & FULL  & 10.604 & 0.505 & 0.103 & 2.983 & 0.497 & 0.055\\
			& CDOE   & 37.334 & 0.513 & 0.193 & 8.367 & 0.500 & 0.092\\ 
			& CCE  & 24.812 & 0.513 & 0.157 & 6.182 & 0.497 & 0.079\\
			& CCPE  & 25.099 & 0.513 & 0.158 & 6.212 & 0.497 & 0.079\\
			& MICE  & 26.262 & 0.484 & 0.161 & 7.170 & 0.481 & 0.083\\
			$\beta_{2}$ & FULL  & 3.900 & 0.198 & 0.062 & 0.981 & 0.201 & 0.031\\
			& CDOE   & 8.256 & 0.196 & 0.131 & 3.690 & 0.200 & 0.061\\ 
			& CCE  & 8.161 & 0.196 & 0.090 & 1.921 & 0.202 & 0.044\\
			& CCPE  & 8.256 & 0.196 & 0.091 & 1.919 & 0.202 & 0.044\\
			& MICE  & 6.942 & 0.195 & 0.083 & 1.747 & 0.198 & 0.042\\\
			$\beta_{3}$ & FULL  & 0.051 & 0.200 & 0.007 & 0.042 & 0.200 & 0.007\\
			& CDOE   & 0.596 & 0.199 & 0.024 & 0.162 & 0.201 & 0.013\\ 
			& CCE  & 0.288 & 0.199 & 0.017 & 0.099 & 0.200 & 0.009\\
			& CCPE  & 0.288 & 0.199 & 0.017 & 0.099 & 0.200 & 0.009\\
			& MICE  & 0.689 & 0.181 & 0.019 & 0.455 & 0.181 & 0.011\\
			$\beta_{4}$ & FULL  & 0.066 & 0.200 & 0.008 & 0.047 & 0.200 & 0.007\\
			& CDOE  & 1.127 & 0.200 & 0.036 & 0.251 & 0.199 & 0.016\\ 
			& CCE  & 0.429 & 0.199 & 0.021 & 0.133 & 0.200 & 0.011\\
			& CCPE  & 0.431 & 0.199 & 0.021 & 0.133 & 0.200 & 0.011\\
			& MICE  & 3.253 & 0.157 & 0.038 & 1.678 & 0.164 & 0.020\\
		\end{tabular}
	}
	\caption{Estimated Mean, Standard Deviation (SD), and Mean Squared Error (MSE) of Different Approaches across Parameters and Sample Sizes, with Missing Covariates and Responses.}
	\label{sim_Res_response}
\end{table}

As expected, FULL has the lowest MSE across all sample size settings and for all $\beta$. Among the four estimators that handle the missing data, similar to the missing covariate-only scenario, MICE has biased estimators across all sample sizes for the intercept, $\beta_1$, $\beta_3$, and $\beta_4$.  CDOE, CCE, and CCPE produce unbiased estimators for all model parameters. Moreover, both CCE and CCPE demonstrate smaller MSE and smaller SD compared to CDOE. However, there is no significant difference in terms of MSE and SD between CCPE and CCE. When comparing MICE to the other missing data handling methods in estimating coefficients of covariates without missing values, different from the results in Section 4.1, now the MSE of MICE is the smallest only for estimating $\beta_2$. For all of the other parameters, CCE and CCPE have the smallest SD and MSE among all estimators excluding FULL.

The most important take-home message of this simulation study is that there is no observed gain of CCPE over CCE. This is consistent with our theoretical explanation in Section 3.
\section{Real-data examples}
\hspace{4mm}
We discuss two real-data examples in this section. The first is regarding a study on wellness assessment of home-care workers, which involves fitting a LMM. The second is related to a study on health policy for end stage renal disease, which involves fitting a logistic regression model. Substantial missing values are involved in both examples.
\subsection{Wellness assessment of home-care workers}
\hspace{4mm}
The data considered in this analysis is part of data collected in a daily study conducted by Dr. Ryan Olson and his colleagues at the Oregon Health and Science University, USA to measure home-care worker's exposure levels to demanding tasks, and also to understand how work exposure might relate to daily physical symptoms, wellness, and lifestyle behaviors.

Twenty-three home-care workers, predominantly females, participated in the study. The participants were 56\% minority, including African American (5), Asian (2), Native American (2), Hispanic/Latino (2), and other (2). The participants worked roughly 30 hours/week, and most (17) regularly performed clients lifting or transferring tasks. After undertaking a baseline demographic and psychosocial survey, the participants completed up to 16 days of daily self-monitoring using the safety task assessment tool developed by the researchers. It was noted that the participants might meet different clients on those days. The work task survey focuses on measuring time and frequency exposures to demanding tasks, and was completed by the worker during the natural breaks at work. The end-of-day survey focuses on recording physical, behavior, and
psychosocial symptoms that could be affected by the work task exposures,
and was completed in the evening between dinner and bedtime.

The analysis here focus on one of the end-of-day survey variables,
namely, the Daily Occupational Fatigue (DOF). Twelve other variables
are used as predictors, including Time Wheelchair (TW; time spent moving
the client around while the client is on wheelchair), Bathing/Toileting
Time (BTT; time spent helping the client bathing and using the toilet),
Dressing Time (DT; time spent on changing the client's clothes),
Transfer Time (TT; time spent completing the transfers), House Clean Time
(HCT; time spent cleaning the client's house), Heavy Object Count (HOC;
number of heavy objects moved; $>$35 lbs/object), Heavy Object Time
(HOT; time spent on moving the heavy objects), Sleep Lag (SL), Pain Lag
(PL), Daily Perceived Stress (DPS), Daily Bad Diet Sum (DBDS), and
Non-lagged Pain (NLP). The data set involves a large number of missing
records, including missing covariates and missing responses.

The following LMM is considered in analyzing the data:
\begin{eqnarray}
	y_{it}&=&x_{it}'\beta+v_{i}+e_{it},
\end{eqnarray}
$i=1,\dots,m, t=1,\dots,n_{i}$, where $i$ represents the worker and
$t$ the study day; $m=23$ is the number of home-care workers; $n_{i}$ is the number of study days for worker $i$; $x_{it}$ is the vector of predictor variables as described above, and $\beta$ is a vector of unknown fixed effects, including an intercept. Furthermore, $v_{i}$ is a worker-specific random effect, and $e_{it}$ is an additional error. It is assumed that the $v_{i}$'s and $e_{it}$'s are independent with $v_{i}\sim N(0,\sigma_{v}^{2})$ and $e_{it}\sim N(0,\sigma_{e}^{2})$.

\begin{table}[htp]
	\centering
	\resizebox{0.9\textwidth}{!}{
		\begin{tabular}{r|rr|rr|rr|rr}
			&\multicolumn{2}{c|}{CDOE} &\multicolumn{2}{c|}{ CCE}&\multicolumn{2}{c|}{ CCPE} &\multicolumn{2}{c}{ MICE}\\
			& coefficient & P-value& coefficient & P-value& coefficient & P-value& coefficient & P-value\\
			\hline
			Intercept & 6.115(3.981) &0.124 & 6.126(3.557) & 0.087& 6.427(2.175) &0.003& 1.459(3.385)& 0.666\\ 
			TW & 0.197(0.251) &0.432 & 0.139(0.090) & 0.123 &0.130(0.029) & $<$0.001& 0.061(0.099)& 0.538\\ 
			BTT & 0.010(0.052) & 0.845 & -0.049(0.039) & 0.215&-0.059(0.012) &$<$0.001 &  -0.054(0.040)&0.177 \\ 
			DT & -0.064(0.069) & 0.358 &0.048(0.035) & 0.168& 0.069(0.013) &$<$0.001& 0.012(0.035)& 0.732 \\ 
			TT & -0.066(0.103) & 0.523& -0.066(0.070) & 0.348& -0.073(0.030) & 0.015&  -0.021(0.078) & 0.788\\ 
			HCT & 0.012(0.009) & 0.217 &0.014(0.007) & 0.090& 0.013(0.003) & $<$0.001& 0.013(0.010) & 0.194 \\ 
			HOC & -0.303(0.594) & 0.611 &-0.267(0.449) & 0.552& -0.290(0.200) & 0.149& -0.015(0.492) & 0.976\\ 
			HOT & 0.011(0.057) & 0.848 &0.014(0.056) & 0.801& 0.007(0.022) & 0.764& 0.036(0.062) & 0.561\\ 
			SL & 0.654(0.506) & 0.197 &0.873(0.418) & 0.039& 0.971(0.136) & $<$0.001& 0.499(0.540) &  0.355\\ 
			PL & 0.169(0.286) & 0.556& 0.021(0.269) & 0.939& -0.057(0.126) & 0.650 & 0.299(0.300) & 0.319\\ 
			DPS & 0.821(0.224) & $<$0.001& 0.672(0.189) & 0.001& 0.506(0.067) & $<$0.001& 0.996(0.226) & $<$0.001 \\ 
			DBDS & 0.221(0.326) & 0.498 &0.409(0.295) & 0.167& 0.512(0.128) &$<$0.001& 0.277(0.292) &  0.343\\ 
			NLP & 0.768(0.281) & 0.006 & 1.001(0.240) & $<$0.001& 1.237(0.119) &$<$0.001& 0.960(0.295) & 0.001\\
		\end{tabular}
	}
	\caption{Analysis of Home-care Workers Data: Estimated Fixed Effects and the Corresponding Standard Errors (in the Parentheses) by Different Methods of Handling the Missing Data}
	\label{real_data_1}
\end{table}

Table \ref{real_data_1} presents the values of parameter estimates, with the standard errors (SEs) in the parentheses, for four different estimators, CDOE, CCE, CCPE and MICE. It is seen that CDOE, CCE and CCPE are mostly similar in terms of the point estimates, while the MICE estimates appear to be quite different, especially for the intercept. Unlike in the simulation study, here we do not know the true values of the parameters. Nevertheless, it can be seen that the SEs of CCE are uniformly smaller than the SEs of CDOE, and mostly smaller than the SEs of MICE. This seems to be consistent with our simulation results in Section 4.1 and 4.3, which suggest that CCE is more efficient than CDOE, and more efficient than MICE for estimating the parameters of the covariates that have missing values.

It should be noted that the reduced SEs do make differences, in some cases, in terms of the statistical significance. For example, at the 10\% level of significance, 5 covariates were found by CCE to be significantly associated with the outcome, while 2 covariates were found either by CDOE or by MICE to be significantly associated with the outcome. The two significant covariates found by CDOE or MICE were also found to be significant by CCE, which are DPS and NLP. In addition, CCE also found the intercept, HCT and SL to be significant. The finding by CCE that the intercept is nonzero is not surprising: Many factors, other than the work, may affect a person's life that, just because no work is involved, the person is not necessarily stress-free.

One ``surprising'' observation is that the SEs of CCPE are much smaller than the SEs of all of the other estimates; as a result, CCPE found a lot more predictors (all except 3) that are significant at 10\% (in fact, 5\%) level. This seems to be inconsistent with our simulation results in Section 4.3, which suggest that CCE and CCPE are almost the same in terms of estimation efficiency; it is also contradicting to our theoretical explanation (see Section 3), which suggests that CCPE is not superior than the CCE. But the SE of CCPE is misleading. Note that, in our simulation study, we reported the (simulated) SD (standard deviation), not the SE (standard error), which means an estimated SD. Unlike the simulated SD in the simulation study, the SE can be a poor estimate of the SD, if something is wrong, and the computer does not know it. This seems to be the case here. Recall the CCPE is based on replacing the missing $y_{\rm m}$ by $\hat{y}_{\rm m}$, the EBP of $y_{\rm m}$ based on $y_{\rm o}$ and $X$ (see Section 3). But the computer could tell the difference between the two, and treats $\hat{y}_{\rm m}$ as observed. This serious ``misunderstanding'' leads to the fake SEs associated with CCPE, which are significantly smaller than the SEs of CCE (and the other estimates), and the false statistical significance as a result.

A take-home message is that CCE is recommended over CCPE, as the latter can produce misleading standard errors, which can impact the inferential results, including false statistical significance and invalid lengths of confidence intervals.
\subsection{Health policy study for ESRD}
\hspace{4mm}
The data involved in this subsection is a sample of 5\% Chronic Kidney Disease data which came from the United States Renal Data System (USRDS; https://www.niddk.nih.gov/about-niddk/strategic-plans-reports/usrds). End-stage renal disease (ESRD) is a major health problem that has led to a significant number of death in the U.S. (e.g., Yang {\it et al.} 2007, USRDS 2023). Potential risk factors include diabetes status, race, age, sex, and other comorbidity (e.g., Hsu {\it et al.} 2001, Ofsthun {\it et al.} 2003, Harris and Zhang 2020). Existing approaches to reduce the mortality rates include early and accurate diagnosis, managing diabetes and high blood pressure, and dialysis. These strategies require accessibility to quality health care, which many U.S. citizens are lacking. On March 23, 2010, President Barack Obama signed the Affordable Care Act (ACA; also known as the Obama Care) into law, which was designed to make affordable health insurance available to more people and to expand Medicaid to cover all adults with income below 138\% of the federal poverty level. It is hypothesized that better health care access will decrease the mortality rate due to ESRD. To test this hypothesis, we examined data from the United States Renal Data System (USRDS) collected in 2008-2010 (before the ACA) and compared it to data collected in 2011-2013 (after the ACA) to see if there was a reduction in mortality due to ESRD. We also examined the impact, if any, of the risk factors of diabetes status, race, age, sex, and other comorbidities such as hematocrit level and hemoglobin level on this relationship. The subjects involved were patients with ESRD. 

Our analysis is based on a logistic regression model. The outcome variable is a binary indicator of death. The predictors (or independent variables) include {\it cohort}, a group indicator for the before (0) or after (1) ACA; {\it diabetes}, an indicator variable on whether the patient had diabetes; {\it hcrit\_Max}, a continuous variable of the maximum hematocrit level from 3-year annual reports for each cohort; {\it hgb\_Max}, another continuous variable of the maximum hemoglobin level from 3-year annual reports for each cohort; {\it SEXCODE}, an indicator variable for male; {\it distinct\_code\_count}, the total number of unique comorbidities; {\it age\_at\_esrd}, the age when a patient developed ESRD; {\it RACECODE}, a multi-categorial variable with seven categories--Native American (NA), Asian (AS), Black (BLK), White (served as the reference level for creating 6-dummy variables), Unknown (UKN), Hispanic (HIS), and Others; and dialysis modality, categorized into seven categories--hemodialysis (HD) Only served as the reference level, HD to Peritoneal (PD) Late Switch, HD to PD Switch, Multi Switch, PD to HD Late Switch, PD to HD Switch, and PD Only. Here, Switch is defined as a switch from HD to PD or vice versa more than six months before the end of their cohort.  Late Switch is defined as a switch less than six months before the end of their cohort.

The data contains a large amount of missing covariates, although there is no missing value in the outcome variable (which makes sense because it was an indication of death or not). \textcolor{red}{T}he variable {\it hgb\_Max} has 71.39\% missing, and the variable {\it hcrit\_Max} has 15.93\% missing. We carried out the statistical analyses using three different approaches of handling the missing values, CDOE, CCE and MICE. The analysis results are presented in Table \ref{real_data_2}. It is seen that all three methods have found {\it cohort} to be a significant factor, confirming that ACA made a difference in reducing the ESRD mortality. To see the difference in terms of the statistical significance of the other predictors, note that, compared to the home-care workers data discussed in Section 5.1, the current data size is much bigger---there were 18,384 observed outcomes. Thus, here, we consider a higher level of significance at 5\% instead of 10\%. The status of statistical significance at the 5\% level (with the $\surd$ sign) as well as the total number of significant logistic regression coefficients (bottom row), are also included in Table \ref{real_data_2}. It is seen that both CCE and MICE have resulted in 12 significant results, while CDOE has resulted in 7.
\begin{table}[ht]
	\centering
	\resizebox{0.9\textwidth}{!}{
		\begin{tabular}{r|lr|lr|lr}
			&\multicolumn{2}{c|}{CDOE} &\multicolumn{2}{c|}{ CCE}&\multicolumn{2}{c}{ MICE}\\
			& coefficient & P-value& coefficient & P-value& coefficient & P-value\\
			\hline
			Intercept & -3.316(0.255) $\surd$ & $<$ 0.001 &-3.083(0.124) $\surd$ & $<$ 0.001 &-2.396(0.118) $\surd$& $<$ 0.001\\ 
			cohort & -1.243(0.088) $\surd$ & $<$0.001 &-1.386(0.036) $\surd$ & $<$ 0.001 & -1.350(0.035) $\surd$& $<$ 0.001 \\ 
			diabetes & 0.030(0.093) & 0.749 & 0.057(0.039) & 0.143 &0.047(0.039)& 0.179 \\ 
			hcrit\_Max & -0.001(0.001) & 0.220 & -0.005(0.001) $\surd$ & $<$ 0.001 & -0.008(0.001) $\surd$&$<$ 0.001 \\ 
			hgb\_Max & -0.000(0.001) & 0.738 & -0.006(0.002) $\surd$ & 0.008& -0.001(0.002)&0.617 \\ 
			SEXCODE & 0.157(0.079) $\surd$ & 0.046 & 0.059(0.034) & 0.081& 0.056(0.034)&0.099\\ 
			distinct\_code\_count & 0.022(0.010) $\surd$ & 0.033 &-0.005(0.004) & 0.276 & -0.019(0.004) $\surd$&$<$ 0.001 \\ 
			age\_at\_esrd & 0.044(0.003) $\surd$ & $<$ 0.001 &  0.051(0.001) $\surd$ & $<$ 0.001 & 0.052(0.001) $\surd$&$<$ 0.001 \\ 
			RACE(NA) & -0.376(0.343) & 0.273 & -0.465(0.159) $\surd$ & 0.003& -0.483(0.158) $\surd$&0.002 \\ 
			RACE(AS) & -0.277(0.229) & 0.227 &-0.869(0.106) $\surd$ & $<$ 0.001& -0.879(0.106) $\surd$& $<$ 0.001 \\ 
			RACE(BLK) & -0.494(0.092) $\surd$ & $<$ 0.001 & -0.403(0.039) $\surd$ & $<$ 0.001& -0.415(0.039) $\surd$&$<$ 0.001 \\ 
			RACE(UKN) & -0.140(0.490) & 0.775 & -0.770(0.253) $\surd$ & 0.002& -0.801(0.251) $\surd$&0.002 \\ 
			RACE(HIS) & -0.713(0.202) $\surd$ & $<$0.001 & -0.798(0.080) $\surd$ & $<$ 0.001 &-0.788(0.080) $\surd$&$<$ 0.001 \\ 
			RACE(Other) & -0.498(0.275) & 0.070 & -0.433(0.122) $\surd$ & $<$ 0.001& -0.461(0.122) $\surd$&$<$ 0.001 \\ 
			HD\_to\_PD\_Late\_Switch & -0.003(0.403) & 0.994 & -0.014(0.205) & 0.946& -0.055(0.204)&0.787 \\ 
			HD\_to\_PD\_Switch & -0.687(0.411) & 0.094 & -0.230(0.154) & 0.135& -0.207(0.153)&0.176 \\ 
			Multi\_Switch & -0.245(0.165) & 0.137 & -0.062(0.082) & 0.448 & -0.083(0.082)&0.311 \\ 
			PD\_to\_HD\_Late\_Switch & 0.165(0.478) & 0.730 & -0.358(0.232) & 0.123& -0.346(0.232)&0.136 \\ 
			PD\_to\_HD\_Switch & 0.046(0.319) & 0.884 & -0.176(0.172) & 0.306 & -0.216(0.172)&0.210 \\ 
			Peritoneal\_Only & -0.055(0.264) & 0.835 & -0.743(0.104) $\surd$ & $<$ 0.001& -0.708(0.104) $\surd$&$<$ 0.001 \\
			Total \# of Sig. (5\%) &\multicolumn{2}{c|}{7} &\multicolumn{2}{c|}{12}&\multicolumn{2}{c}{ 12}
		\end{tabular}
	}
	\caption{Analysis of ESRD Data: Estimated Coefficients of Logistic Regression (Standard Errors in the Parentheses) by Different Methods of Handling the Missing Data, as well as Status ($\surd$ Indicating Yes) and Total \# (Bottom Row) of Statistical Significance at 5\% Level.}
	\label{real_data_2}
\end{table}

As for the comparison between CCE and MICE, although both have produced 12 statistically significant results, they are different in two of the corresponding variables, namely, {\it hgb\_Max} and {\it distinct\_code\_count}. The CCE result is significant for {\it hgb\_Max} but not for {\it distinct\_code\_count}; the MICE result is the other way around. Note that there is a huge difference in terms of the percentage of missing data for these two variables. For {\it hgb\_Max}, 71.39\% are missing; for {\it distinct\_code\_count}, only 0.24\% are missing. According to the results of our simulation study (see Section 4.2), CCE is more efficient for variables with a larger percentage of missing values while MICE is just the opposite, more efficient for variables with no missing values. This seems to be consistent with the findings of this real-data analysis.
\section{Discussion}
\hspace{4mm}
The proposed random effects approach to missing data is a natural way of handling missing values in the GLMM analysis, because there are already random effects in the model anyway. We showed that, as long as a response is observed, the missing covariates can be mended into a GLMM with no missing covariates (the CCE approach), based on which parameter estimation can be done more efficiently compared to the complete-records-only analysis.

On the other hand, we showed that there is nothing one could do for the missing responses, even if the latter are replaced by their empirically best predictor based on the observed data (the CCPE approach). In fact, we have explored another approach, in which the missing responses are substituted by bootstrapped data sampled parametrically based on parameters estimated by the observed data. Once again, our simulation results showed no gain over the CCE.

This raises a question on whether CCE is indeed the best one can do. We believe it is as long as the analysis is based on a GLMM, which is a model conditional on the covariates, $X$, and all the inferential measures, especially the variance, are also conditional on $X$. Recall the notation $y_{\rm o}$, $y_{\rm m}$, $X_{\rm o}$ and $X_{\rm m}$ (see Section 3). The CCE is based on completing $X_{\rm o}$, and fitting the LMM, (\ref{eq:sub_LMM_o}). On the other hand, $y_{\rm m}$ is missing, but what about $X_{\rm m}$? There may still be observed components in $X_{\rm m}$, which could provide additional information. Our take on this is that, as long as the estimation of $\beta$ is concerned, the useful information should be regarding the relationship between $X$ and $y$. If the $y$ part is missing, the $X$ part is not relevant even if it is observed. However, (the observed part of) $X_{\rm m}$ can still provide additional information if a marginal model is considered, in which both $X$ and $y$ are considered random observations, and the inferential measures, such as the variance, is unconditional, taking into account the variation in $X$.

The proposed random-effects approach is a ``missing at random (MAR)'' scenario (although not necessarily ``missing completely at random''). Note that existing imputation-based methods, such as MICE, are also based on the MAR framework.

Our empirical studies suggest the advantage of the CCE over MICE is more significant under an LMM than under a logistic regression model. We conjecture that the same is also generally true for a GLMM that is not an LMM, and there is an explanation for this. It is known (e.g., Jiang and Nguyen 2009) that, compared to LMM, GLMM is more sensitive to violation of the distributional assumption about the random effects. In CCE, additional random effects are introduced to incorporate the missing values. Such random effects may not satisfy the assumed distributional assumption. A LMM is more robust to such a violation of the distributional assumption than a GLMM; such robustness may affect the efficiency of the CCE. As for the comparison between MICE and CCE in the case of GLMM, our empirical results suggest that MICE is relatively more efficient for estimating coefficients of the covariates that have no or small percentages of missing values, while CCE is more efficient for estimating coefficients of the covariates that have large percentages of missing values. In a way, the latter is more important as the covariates with large percentages of missingness are the ones who are more in need for help.

Finally, as illustrated by our real-data example, the CCPE method can produce false statistical significance. See discussions in the last two paragraphs of Section 5.1. Note that, in this case, the term ``computer'' refers to the software package used to perform the statistical analysis. There is a difference between a software package and an artificial intelligence (AI) tool in that the former may be fooled in some ways but the latter is not supposed be, because an AI tool is to work like a human being, who knows what is going on. As quoted from Peter Bickel's 2013 R. A. Fisher Lecture, ``Applied statistics must engage theory and conversely." To be not fooled by the output of a software package requires engaging the theory, as we have done in this work (see Sections 2 and 3), in addition to knowing how to run the software.

\vspace{5mm}

{\bf Acknowledgements.} Thuan Nguyen and Jiming Jiang's research is partially supported, respectively, by the NSF grants DMS-2210372 and DMS-2210569. The data for the health policy study were obtained under the support of Thuan Nguyen's NSF grant SES-1118469. The authors are grateful to Drs. Ryan Olson and Brad Wipfli of the Center for Research on Occupational and Environmental Toxicology at the Oregon Health and Science University for providing the data on the home-care worker wellness assessment from their research as well as information and helpful discussions.
\bibliographystyle{biometrika}

\begin{thebibliography}{99}
	\bibitem{}
	Afifi, A. and Elashoff, R. (1966), Missing observations in multivariate
	statistics: I. Review of the literature, {\it J. Amer. Statist. Assoc.} 61, 595--604.
	\bibitem{}
	Datta, G. S. \& Lahiri, P. (2000), A unified measure of
	uncertainty of estimated best linear unbiased predictors
	in small area estimation problems, {\it Statist. Sinica} 10, 613--627.
	\bibitem{}
	Diggle, P. J., Heagerty, P., Liang, K. Y., and Zeger, S. L. (2002),
	{\it Analysis of Longitudinal Data}, 2nd ed., Oxford Univ. Press.
	\bibitem{}
	Harris, R. C. and Zhang, M.-Z (2020), The Role of Gender Disparities in Kidney Injury, {\it Ann. Transl. Med.} 8, 514--514.
	\bibitem{}
	Hartley, H. O. and Hocking, R. (1971), The analysis of incomplete data, {\it Biometrics} 27, 783--823.
	\bibitem{}
	Hsu, C.-., et al. (2001), Relationship between Hematocrit and Renal Function in Men and Women, {\it Kidney Int.} 59, 725--731.
	\bibitem{}
	Jiang, J. (1998), Asymptotic properties of the empirical BLUP and BLUE in mixed linear models, {\it Stat. Sin.} 8, 861--885.
	\bibitem{}
	Jiang, J. (2022), {\it Large Sample Techniques for Statistics}, 2nd ed., Springer, New York.
	\bibitem{}
	Jiang, J. and Nguyen, T. (2009), Comments on Goodness-of-fit
	tests in mixed models by G. Claeskens and J. D. Hart, {\it TEST} 18,
	248--255.
	\bibitem{}
	Jiang, J. and Nguyen, T. (2021), {\it Linear and Generalized Linear Mixed Models and Their Applications}, 2nd ed., Springer, New York.
	\bibitem{}
	Jiang, J., Wand, M. P. and Bhaskaran, A. (2022), Usable and
	precise asymptotics for generalized linear mixed model analysis and
	design, {\it J. Roy. Statist. Soc., B} 84, 55--82.
	\bibitem{}
	Laird, N. M. \& Ware, J. M. (1982), Random effects models
	for longitudinal data, {\it Biometrics} 38, 963--974.
	\bibitem{}
	Little, R. J. A. and Rubin, D. B. (2014), {\it Statistical Analysis
		with Missing Data}, 2nd ed., Wiley, Hoboken, NJ.
	\bibitem{}
	Liu, J. S. (2004), {\it Monte Carlo Strategies in Scientific Computing}, Springer, New York.
	\bibitem{}
	Lu, T.-T. and Shiou, S.-H. (2002), Inverses of $2\times 2$ block matrices, {\it Comput. Math. Appl.} 43, 119--129.
	\bibitem{}
	McCullagh, P. and Nelder, J. A. (1989). {\it Generalized Linear Models},
	2nd ed., Chapman and Hall, London.
	\bibitem{}
	Ofsthun, N., Labrecque, J., Lacson, E., Keen, M., and Lazarus, J. M. (2003), The effects of higher hemoglobin levels on mortality and hospitalization in HD patients, {\it Kidney int.} 63, 1908--1914.
	\bibitem{}
	Rubin, D. B. (1976), Inference and missing data, {\it Biometrika} 63, 581--592.
	\bibitem{}
	United States Renal Data System (2023), {\it USRDS dataset: 2023 data release}, National Institutes of Health, National Institute of Diabetes and Digestive and Kidney Diseases, Bethesda, MD.
	\bibitem{}
	Van Buuren, S. and Groothuis Oudshoon, K. (2011),  MICE: Multivariate
	Imputation by Chained Equations in R, {\it J. Stat. Softw.} 45, 1--67.
	\bibitem{}
	White, H. (1982), Maximum likelihood estimation of misspecified models, {\it Econometrica} 50, 1--25.
	\bibitem{}
	Yang, W., Israni, R. K., Brunelli, S. M., Joffe, M. M., Fishbane, S., and Feldman, H. I. (2007), Hemoglobin variability and mortality in ESRD, {\it J. Am. Soc. Nephrol.} 18, 3164--3170.
\end{thebibliography}

\end{document}